\documentclass[twocolumn]{aastex701}

\usepackage{graphicx}
\usepackage{multirow}
\usepackage{url}
\usepackage{color, hyperref, epsfig}
\usepackage{apjfonts, natbib}
\usepackage{appendix}
\usepackage{amsmath}
\usepackage{float}
\usepackage{bm}
\usepackage{booktabs}
\usepackage{ragged2e}

\usepackage[utf8]{inputenc}
\DeclareUnicodeCharacter{02BC}{'} %

\def\s{{\rm\thinspace s}}

\newcommand{\src}{{\rm\, J1105+1452 }}

\begin{document}



\title{A Radio Changing-state Jet in the Narrow-line Seyfert 1 Galaxy J1105+1452}

\author[0000-0002-4757-8622]{Liming Dou}
\affiliation{Department of Astronomy, Guangzhou University, Guangzhou 510006, People's Republic of China; doulm@gzhu.edu.cn}
\email{doulm@gzhu.edu.cn}

\author{Zhining Chen}
\affiliation{Department of Astronomy, Guangzhou University, Guangzhou 510006, People's Republic of China; doulm@gzhu.edu.cn}
\email{2112419037@e.gzhu.edu.cn}

\author{Jiahua Wu}
\affiliation{Department of Astronomy, Guangzhou University, Guangzhou 510006, People's Republic of China; doulm@gzhu.edu.cn}
\email{jhwu@e.gzhu.edu.cn}

\author{Ning Jiang}
\affiliation{Department of Astronomy, University of Science and Technology of China, Hefei, 230026, People’s Republic of China}
\email{jnac@ustc.edu.cn}

\author{Xinwen Shu}
\affiliation{Department of Physics, Anhui Normal University, Wuhu, Anhui 241002, People’s Republic of China}
\email{xwshu@ahnu.edu.cn}

\author{Tinggui Wang}
\affiliation{Department of Astronomy, University of Science and Technology of China, Hefei, 230026, People’s Republic of China}
\email{twang@ustc.edu.cn}

\author{Xiaofeng Li}
\affiliation{School of Computer Science and Information Engineering, Changzhou Institute of Technology, Changzhou, Jiangsu 213032, People's Republic of China}
\email{lixiaofeng@czu.cn}

\author{Gege Wang}
\affiliation{College of Physical Science and Technology, Shenyang Normal University, Shenyang 110034, Peopleʼs Republic of China}
\email{wanggege@synu.edu.cn}

\author{Yanli Ai}
\affiliation{College of Engineering Physics, Shenzhen Technology University, Shenzhen 518118, People's Republic of China; aiyanli@sztu.edu.cn}
\email{aiyanli@sztu.edu.cn}
 
\author{Mengzhu Chen}
\affiliation{Department of Physics, Guangzhou University, Guangzhou 510006, People's Republic of China} 
\email{cmz1104@e.gzhu.edu.cn}
 
\author{Jianguo Wang}
\affiliation{Key Laboratory for the Structure and Evolution of Celestial Objects, Yunnan Observatories, Kunming 650011, People's Republic of China}
\email{wangjg@ynao.ac.cn}

\author{Junhui Fan}
\affiliation{Department of Astronomy, Guangzhou University, Guangzhou 510006, People's Republic of China; doulm@gzhu.edu.cn}
\email{fjh@gzhu.edu.cn}

\begin{abstract}

We report the discovery of a radio-quiet to radio-loud transition in the narrow-line Seyfert~1 galaxy J1105+1452. The source has undergone a long-term evolution from a radio-quiet state in the 1990s to a persistently radio-bright state after 2017. Post-2017 flux densities in the $0.8$--$7$~GHz range cluster between $32$ and $43$~mJy, whereas the $144$~MHz flux density is only $1.94 \pm 0.23$~mJy. This indicates strong low-frequency suppression from a compact, absorbed component. Modeling the radio spectral energy distribution with a synchrotron self-absorption model yields a turnover frequency $\nu_{\rm p} = 0.48 \pm 0.03$~GHz and a peak flux density $S_{\rm p} = 38.9 \pm 4.7$~mJy. These parameters classify J1105+1452 as a megahertz peaked-spectrum source, consistent with the new episode of an early-stage compact jet. Under the assumption of equipartition, we derive an intrinsic physical radius $R \sim 0.68$\,pc and an average apparent expansion velocity $\beta_{\rm app} \approx 0.64$. The observed brightness temperature $T_b \approx 6.0 \times 10^{11}$\,K necessitates a Doppler factor $\delta \approx 12$, implying a relativistic jet viewed at $\theta \lesssim 5^\circ$. Despite the dramatic radio evolution, the X-ray spectrum remains stable and steep ($\Gamma \simeq 3.0$), suggesting that the X-ray emission remains dominated by the disk--corona, while the radio band has become jet-dominated. Our results identify J1105+1452 as a rare radio changing-state NLSy1, providing a unique laboratory for studying the birth and early evolution of relativistic jets at high Eddington ratios.

\end{abstract}


\section{Introduction}\label{sec:intro}

Active galactic nuclei (AGNs) are powered by accretion onto supermassive black holes (SMBHs) and provide a key channel for coupling black hole growth to host-galaxy evolution \citep[e.g.,][]{Kormendy2013}.
Although orientation-based unification explains much of the observed phenomenology \citep[e.g.,][]{Urry1995}, the physical conditions that launch and sustain relativistic jets remain uncertain. Powerful, persistent jets are most commonly associated with blazars hosted by massive ellipticals and are often discussed in the context of low accretion rates \citep[e.g.,][]{Heckman2014}.

Narrow-line Seyfert~1 (NLSy1) galaxies probe jet launching in high-accretion, low-mass SMBH systems. NLSy1s are characterized by narrow permitted lines and strong optical Fe~{\sc ii} emission \citep[e.g.,][]{Osterbrock1985}, and typically accrete at high Eddington ratios \citep[e.g.,][]{Mathur2000}. Although only $\sim$7\% are radio loud \citep[e.g.,][]{Komossa2006}, some exhibit blazar-like properties, including flat radio spectra and GeV emission \citep[e.g.,][]{Yuan2008,Abdo2009}, raising the question of how compact jets emerge in high-accretion environments.

The variability of radio-loud NLSy1s has been studied across a wide range of timescales, from intra-night fluctuations to decadal transitions\citep[e.g.,][]{Jha2026}. Sources with confirmed $\gamma$-ray emission often show high levels of intra-night optical variability, a trait they share with core-dominated blazars. In contrast, non-jetted NLSy1s and radio-quiet quasars typically show stochastic, disk-driven variability with less coherence over long timescales \citep{Sudan2025}.

\citet[][]{2025A&A...702L..17G} recently reported an extreme radio outburst from the NLSy1 SDSS~J110546.07+145202.4 (hereafter, \src) at $z=0.120886$, during which the radio flux density increased by a factor of $\sim$20 and has remained elevated for multiple years after 2017, placing the source in an exceptionally radio-loud state. The reported flat spectrum suggests a compact, core-dominated component, but the physical origin of the event remains unclear.

Following the release of the Very Large Array Sky Survey (VLASS) catalog, we investigated radio-variable sources by correlating their radio flux densities with mid-infrared light curves \citep[e.g.,][]{Dai2020}. Independently, during a systematic search within a sample of NLSy1s \citep[][]{Wu2026}, we identified extreme radio variability in \src. This discovery prompted our Target of Opportunity (ToO) X-ray observations with {\it Swift}/XRT in 2023 \citep{Burrows2005} and subsequent monitoring with the {\it Einstein Probe} ({\it EP}) Follow-up X-ray Telescope (FXT) in 2025 \citep{Yuan2025}.

Here we extend the radio characterization with newly available low-frequency coverage including LoTSS and present coordinated optical, MIR, and X-ray constraints to test whether the event reflects jet emergence, accretion-state changes, or environmental absorption. 
Throughout this work we assume a flat $\Lambda$CDM cosmology with $H_{0}=67.4$~km~s$^{-1}$~Mpc$^{-1}$, $\Omega_{\rm m}=0.315$, and $\Omega_{\Lambda}=0.685$ \citep{2020A&A...641A...6P}.

\section{Data and Analysis} \label{sec:data}

\subsection {Radio data and Analysis}

\begin{figure}[htbp]
\centering
\plotone{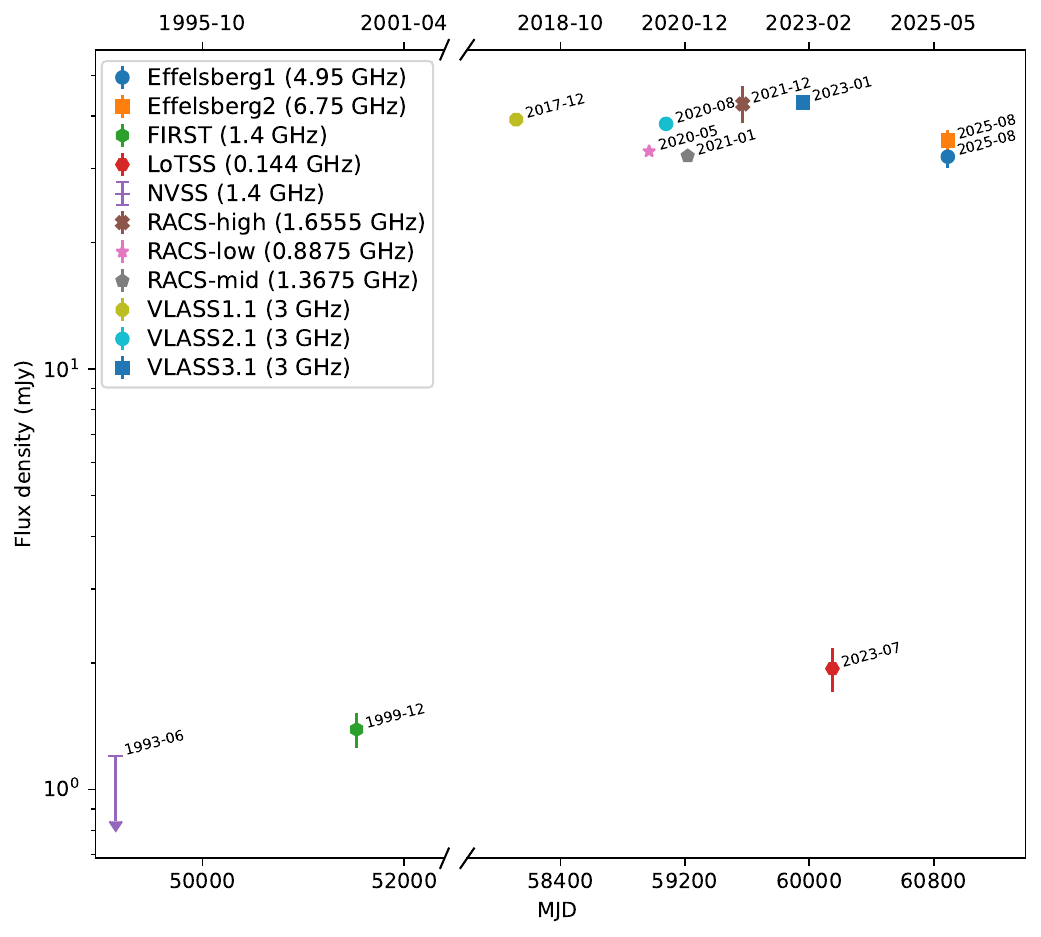}
\caption{Radio light curve of \src\. Data points are aggregated from NVSS, FIRST, LoTSS, RACS (low/mid/high), VLASS (epochs 1.1, 2.1, 3.1), and 4.95 and 6.75\,GHz observations obtained with the Effelsberg 100\,m telescope in 2025 August.
\label{fig:radiolc}}
\end{figure}

We compiled multi-epoch radio flux densities for \src\ from several major sky surveys to characterize its long-term variability. These include the NRAO VLA Sky Survey \citep[NVSS;][]{Condon1998}, the Faint Images of the Radio Sky at Twenty-Centimeters survey \citep[FIRST;][]{Helfand2015}, and the LOFAR Two-metre Sky Survey \citep[LoTSS;][]{Shimwell2022, Varglund2025}. High-cadence and recent measurements were incorporated from the Rapid ASKAP Continuum Survey at low, mid, and high frequencies \citep[RACS;][]{Hale2021, Duchesne2024,Duchesne2025}, as well as multiple epochs from the VLA Sky Survey \citep[VLASS1.1, 2.1, and 3.1;][]{Gordon2021}. Additionally, we included 4.95 and 6.75GHz observations obtained with the Effelsberg 100m telescope in 2025 August \citep{2025A&A...702L..17G}. 
The resulting multi-frequency radio light curve is presented in Figure~\ref{fig:radiolc}. 

\begin{figure}[htbp]
\plotone{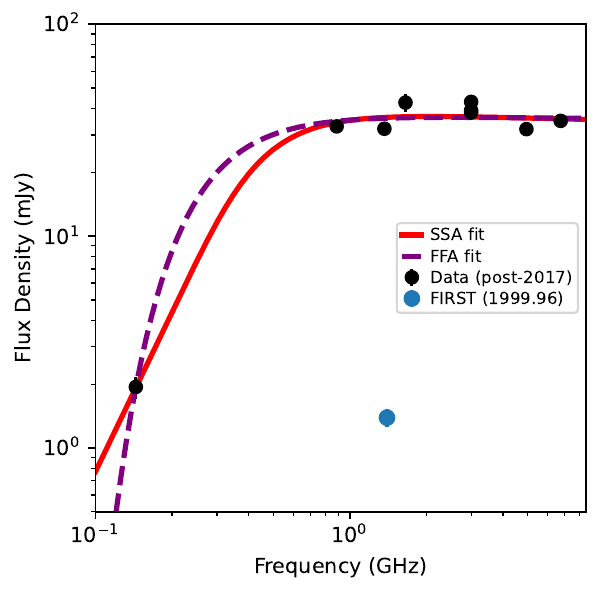}
\caption{Radio spectral energy distribution (SED) of \src\ with best-fit models overlaid. 
\label{fig:radiosed}}
\end{figure}

As illustrated in Figure~\ref{fig:radiolc}, \src\ underwent a dramatic transition from a radio-quiet state (\(F_{\nu}=1.39\pm0.13\)\,mJy) in the 1990s to a persistently bright phase after 2017, with flux densities increasing by a factor of $\gtrsim$20. In recent epochs, the 0.8--7\,GHz flux densities cluster between 30 and 40\,mJy (see Figure~\ref{fig:radiosed}), consistent with a flat radio spectrum ($\alpha\simeq0.1$, where $S_\nu\propto\nu^{-\alpha}$). However, the 144\,MHz flux density from LoTSS is only \(1.94\pm 0.23\)\,mJy. The $\sim$15--20 suppression at 144\,MHz relative to the GHz band requires a spectral turnover and favors a compact, partially self-absorbed component.

We model the radio SED with synchrotron self-absorption (SSA) and free--free absorption (FFA) to characterize the turnover. Assuming a homogeneous spherical source, the SSA flux density \(S_{\nu }\) is given by:
\begin{equation}
S_{\nu} = S_{0} \left( \frac{\nu}{\nu_{p}} \right)^{5/2} \left[ 1 - \exp \left( -\tau_{p} \left( \frac{\nu}{\nu_{p}} \right)^{-(p+4)/2} \right) \right],
\end{equation}
where \(S_{0}\) is the normalization factor, \(\nu _{p}\) is the turnover frequency, \(\tau _{p}\) is the optical depth at \(\nu _{p}\), and \(p\) is the electron energy distribution index. In the optically thin regime (\(\nu \gg \nu _{p}\)), the spectrum follows \(S_{\nu }\propto \nu ^{-\alpha }\), where \(\alpha =(p-1)/2\). In the optically thick regime (\(\nu \ll \nu _{p}\)), the spectrum approaches the characteristic SSA limit of \(S_{\nu }\propto \nu ^{2.5}\). Alternatively, we employ an FFA model where the intrinsic power-law spectrum is attenuated by an external ionized screen: \(S_{\nu }\propto \nu ^{-\alpha }\exp (-\tau _{\nu })\), with \(\tau _{\nu }=(\nu /\nu _{p})^{-2.1}\).

Both models adequately reproduce the SED (Figure~\ref{fig:radiosed}).
Under the SSA framework, we derive the peak flux density \(S_{p}=38.9\pm 4.7\)~mJy at \(\nu _{p}=0.48\pm 0.03\)~GHz. The 144MHz measurement lies deep within the optically thick regime (\(S_{\nu }\propto \nu ^{2.5}\)), and the SSA extrapolation is in excellent agreement with the observed LoTSS flux. This identifies \src\ as a Megahertz Peaked-Spectrum (MPS) source \citep[e.g.,][]{Coppejans2015,Ballieux2024}.

The FFA fit yields an intrinsic optically thin spectral index $\alpha_{\rm thin}=0.01\pm0.06$ and optical depth $\tau_{1\,{\rm GHz}}=0.051\pm0.004$. External free--free absorption is physically plausible in NLSy1s if dense circumnuclear ionized gas (e.g., NLR material) acts as a foreground screen. However, the current lack of measurements below the turnover frequency prevents a robust test of the exponential low-frequency curvature characteristic of FFA. In contrast, SSA provides a natural explanation for the observed low-frequency suppression, while deeper sub-turnover sampling will be required to distinguish between SSA and FFA.

\subsection{X-ray Observations and Analysis} \label{sec:xray}

We observed \src\ with the {\it Swift}/XRT observation in photon-counting mode on 2023 October 30 (UT) with a 2.4~ks exposure. Data were reduced using \texttt{HEASOFT} (v6.33) and the \texttt{xrtpipeline} task. Source counts were extracted from a $47\farcs2$-radius circular aperture centered on the optical position; background was taken from a nearby source-free region. The \(0.3\)--\(10\)keV net count rate was \(0.011\pm 0.002\) counts~s\(^{-1}\), with no significant intra-observation variability.

The {\it EP}/FXT further observed \src\ over four epochs between 2025 January and May. 
Each exposure was $\sim$1.2~ks. FXT-A and FXT-B modules were operated in full-frame mode. Data processing was conducted via \texttt{fxtchain} within the FXT Data Analysis Software (\texttt{FXTDAS}), handling event screening, gain correction, and bad-pixel flagging. Source products were extracted from \(120^{\prime \prime }\) radius regions.

We fit in XSPEC using C-statistics on spectra grouped to a minimum of two counts per bin. 
All uncertainties are reported at the 90\% confidence level. We modeled the spectra with a power law (\(N_{\mathrm{H}}\) fixed at \(1.32\times 10^{20}\)~cm\({}^{-2}\); \citealt{2016A&A...594A.116H}). Fluxes in the \(0.5\)--\(8.0\)~keV band were determined using the \texttt{cflux} model.

The \textit{Swift}/XRT spectrum yielded a photon index of $\Gamma = 2.6 \pm 0.6$ and a $0.5$--$8.0$\,keV flux of $2.6_{-1.0}^{+1.6} \times 10^{-13}$\,erg\,cm$^{-2}$\,s$^{-1}$ ($C/\nu = 10.89/11$). Joint fits for the four \textit{EP}/FXT epochs provided consistent results, with $C/\nu$ values between $24.1/27$ and $25.6/22$. FXT fluxes ranged from $0.7_{-0.5}^{+0.8}$ to $3.1_{-1.3}^{+3.0} \times 10^{-13}$\,erg\,cm$^{-2}$\,s$^{-1}$, with $\Gamma$ spanning $2.4 \pm 0.8$ to $3.6_{-1.4}^{+1.8}$. Additionally, \src\ was detected during the \textit{eROSITA} all-sky survey \citep{Merloni2024} in 2020 May ($0.15$\,ks exposure), yielding $\Gamma = 3.53^{+0.97}_{-0.86}$ and a flux of $1.0_{-0.5}^{+1.0} \times 10^{-13}$\,erg\,cm$^{-2}$\,s$^{-1}$ ($C/\nu = 5.64/7$).

\begin{figure}[htbp]
\centering
\plotone{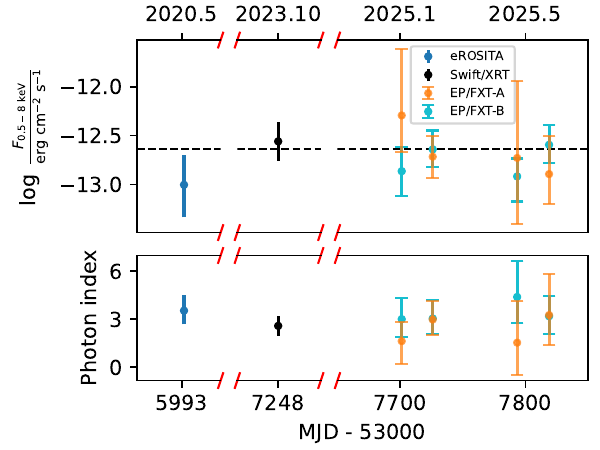}
\caption{Long-term X-ray monitoring of \src\ from 2020 to 2025. 
Top: Observed $0.5$--$8$ keV flux ($\text{erg cm}^{-2}\text{ s}^{-1}$) in log scale. The horizontal dashed line indicates the historical {\it ROSAT} flux level extrapolated to the $0.5$--$8$ keV band assuming a power-law model with $\Gamma = 3.0$.
Bottom: Evolution of the photon index ($\Gamma$) over the same period.
\label{fig:xlc}}
\end{figure}

The long-term X-ray light curve is shown in Figure~\ref{fig:xlc}. A \(\chi^{2}\) test for variability across the {\bf six} 2020--2025 epochs yields \(\chi ^{2}={4.087}\) (dof = {5}, \(p={0.54}\)), indicating the source remained statistically stable with an average \(0.5\)--\(8\)~keV flux of \({1.79}\times 10^{-13}\)erg~cm\({}^{-2}\)~s\({}^{-1}\) and a median \(\Gamma \approx 3.0\).

Historically, \src\ was detected by {\it ROSAT} on November 15, 1990. It was included in the {\it ROSAT} All-Sky Survey Bright Source Catalog \citep{Voges1999} with a 0.1--2.4~keV count rate of 0.09±0.02~counts~s$^{-1}$.
Assuming $\Gamma = {3.0}$, we converted this to a $0.5$--$8$~keV flux of {$\approx 2.3 \times 10^{-13}$~erg~cm$^{-2}$~s$^{-1}$} using \texttt{WebPIMMS}. Within uncertainties, this historical flux is consistent with the current state observed by {\it eROSITA}, {\it Swift} and {\it EP} in 2020--2025 (Figure~\ref{fig:xlc}). This suggests the decades-long radio evolution is not accompanied by a comparably large change in the X-ray continuum.

\subsection{Fermi-LAT Data Analysis} \label{sec:Gammaray}

We analyzed the {\it Fermi} Large Area Telescope (LAT) Pass 8 data for \src, covering the period from 2017 to 2025. Following the standard recommendations from the LAT team, we selected events in the 0.2–500 GeV energy range within a region of interest (ROI) centered on the 4FGL-DR4 position. Data reduction was performed using  \texttt{Fermitools} v2.0.8 with the P8R3\_SOURCE\_V2 instrument response functions. The source model included all 4FGL-DR4 sources within the ROI, with the Galactic (gll\_iem\_v07.fits) and isotropic (iso\_P8R3\_SOURCE\_V2\_v1.txt) backgrounds as free components. Normalization and spectral indices for sources within 5 degree of the target, as well as highly variable sources (variability index; \citealt{ace15}), were left free. 

No significant $\gamma$-ray emission was detected at the position of \src, with a test statistic (TS) derived from the TS map (TS$<$9). Consequently, we calculated a 95\% confidence level upper limit using a binned likelihood analysis. Assuming a power-law spectrum with a photon index of 1.468 (derived from the likelihood fit), the integrated flux upper limit in the 0.2–500 GeV range is $1.18 \times10^{-14}\;{\rm ph}\;{\rm cm}^{-2}\;{\rm s}^{-1}$. This non-detection, combined with the radio flare, constrains the high-energy contribution from the nascent jet.

\subsection {Optical and Infrared Photometry}

\citet{2025A&A...702L..17G} examined the archives of ASAS-SN \citep{Shappee2014} and the Catalina Sky Survey \citep[CSS;][]{Drake2009}, covering observations since 2014 and 2007, respectively. Their analysis revealed no systematic long-term rise or high-amplitude optical outbursts during these periods. 

Historically, \src\ was observed by SDSS on 2003 January 28 (MJD 52667), with \textit{petroMag} magnitudes of $18.35\pm0.04$, $17.53\pm0.01$, $16.97\pm0.01$, $16.62\pm0.01$, and $16.57\pm0.05$ in the $ugriz$ bands, respectively. Using the SDSS $g$ and $r$ bands and the empirical conversion from \citet{Drake2013}, we derived a CSS $V$-band equivalent magnitude of $17.20\pm0.05$. This historical brightness is $\sim$0.4~mag fainter than the average CSS $V$-band photometry ($16.80\pm0.08$~mag) recorded during 2007--2016. However, such a modest discrepancy may not represent an intrinsic AGN flux increase, as it could be attributed to systematic offsets between the two surveys, particularly the varying levels of host galaxy contamination within their respective apertures. 

To more accurately track the core activity, we further consider the SDSS \textit{psfMag} magnitudes, which were $19.06\pm0.02$, $18.65\pm0.02$, $18.15\pm0.02$, $17.75\pm0.02$, and $17.68\pm0.02$ in the $ugriz$ bands. These point-source measurements provide a consistent baseline for comparison with the forced-photometry products from the Asteroid Terrestrial-impact Last Alert System \citep[ATLAS;][]{Tonry2018}, the Zwicky Transient Facility \citep[ZTF;][]{Masci2019}, and Pan-STARRS1 \citep[PS1;][]{Flewelling2020}.

The ATLAS has monitored \src\ in the \(c\) and \(o\) bands since 2016 January. We retrieved forced photometry from difference images via the ATLAS forced photometry server \citep{Shingles2021}. To optimize data quality, we retained only observations with a limiting magnitude \(>17\)~mag and applied weekly binning to enhance the signal-to-noise ratio (S/N). 
Similarly, we obtained forced photometry from the 
ZTF, which was also rebinned weekly for consistency. 
For Pan-STARRS1, 
we queried the PS1 DR2 database at MAST via CasJobs\footnote{https://mastweb.stsci.edu/ps1casjobs}
and extracted $gri$ band forced PSF flux, requiring \texttt{FpsfQfPerfect} >0.9; these measurements were likewise binned into 7-day intervals.

Mid-infrared (MIR) observations were retrieved from the {\it Wide-field Infrared Survey Explorer} \citep[{\it WISE};][]{Wright2010} and the Near-Earth Object {\it WISE} Reactivation mission \citep[NEOWISE;][]{Mainzer2014}. Following the cryogenic and post-cryogenic phases, NEOWISE provided consistent monitoring from 2013 December until its formal conclusion in 2024 August. The {\it WISE} survey sampled the source field at six-month intervals, yielding $\sim$12 exposures within a 24-hour window per epoch. Following \citet{Jiang2021}, we perform image subtraction on co-added WISE/NEOWISE data and measure transient MIR fluxes via PSF photometry on the difference images.

\begin{figure}[htbp]
\centering
\plotone{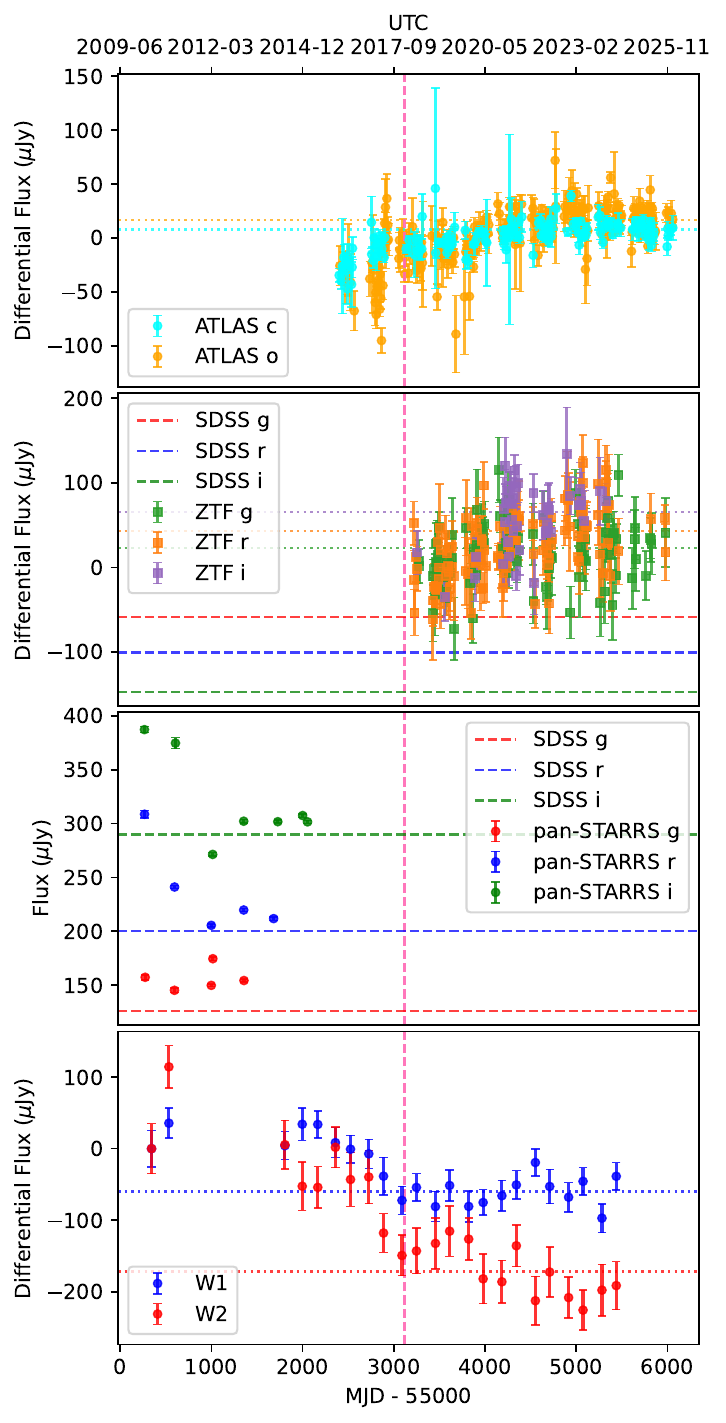}
\caption{Multi-band optical and MIR differential light curves of \src.
Top and Second panels: Difference-flux light curves from ATLAS ($c, o$ bands) and ZTF ($g, r, i$ bands) derived via image subtraction. 
Third panel: Absolute flux measurements from Pan-STARRS1 ($g, r, i$ bands).
Bottom panel: MIR difference-flux light curves ($W1, W2$) from WISE/NEOWISE images. The horizontal dashed lines represent the historical SDSS \textit{psfMag} flux levels from 2003, while the horizontal dotted lines denote the mean flux levels during the post-outburst stabilized state (post-2021 for optical and post-2018 for MIR). 
The vertical dashed line marks the epoch of the 3~GHz radio observation from VLASS1.1.
\label{fig:lc}}
\end{figure}

As illustrated in Figure~\ref{fig:lc}, the MIR light curves exhibit a prominent fading trend prior to 2018. Subsequently, the $W1$-band emission stabilized, while the $W2$ band shows a less distinct transition, likely due to a lower S/N. A qualitatively similar fading-then-stabilization pattern is evident in the Pan-STARRS $r$- and $i$-band light curves between 2009 and 2014. In contrast, both the ATLAS and ZTF light curves exhibit minimal stochastic variability, particularly post-2021. By comparing the early-time data to the mean flux levels of the stabilized state, we identify a gradual rising trend in both ATLAS and ZTF bands prior to 2021. 

Both the PS1 and ZTF measurements show a significant enhancement relative to the historical SDSS \textit{psfMag} baseline. Specifically, the post-2021 ZTF brightness in the $g$, $r$, and $i$ bands is $\sim$0.4~mag brighter than the 2003 SDSS \textit{psfMag} levels. Although this corresponds to a relatively small absolute flux increment of $\sim$0.1--0.2~mJy, the consistent offset across multiple filters and surveys suggests a genuine, albeit modest, long-term increase in the nuclear activity.

\section{Discussion}\label{sec:discussion}

Historically, \src\ was classified as a radio-quiet NLSy1 by \citet{Paliya2024} with a radio-loudness of $R_{\mathrm{5\,GHz}} \approx 4.6$, based on 1.4~GHz FIRST data and 2003 SDSS \textit{cMag} photometry. To characterize its current state, we re-evaluated $R_{\mathrm{5\,GHz}}$ using contemporary multi-wavelength data. Adopting the 5~GHz flux density from our SSA model and post-2021 ZTF $g$- and $r$-band measurements, we find $R_{\mathrm{5\,GHz}} \approx 173$, marking a transition into an extremely radio-loud state. For consistency, we re-calculated the archival radio-loudness using 1993 FIRST flux and 2003 SDSS \textit{psfMag} photometry, yielding $R_{\mathrm{5\,GHz}} \approx 9.6$. This confirms that while \src\ was historically radio-quiet, its current radio activity has intensified by a factor of $\sim$18.

Our SSA model implies a post-2017 1.4~GHz flux density of $\approx$35.9~mJy, corresponding to a luminosity of $L_{1.4\,\mathrm{GHz}} \approx 1.25 \times 10^{24}$~W~Hz$^{-1}$. This significantly exceeds the $\sim$$10^{23}$~W~Hz$^{-1}$ threshold where star formation typically dominates \citep[e.g.,][]{Gurkan2018}, confirming that the radio emission is AGN-powered. 

While the FIRST-era emission might originate from a pre-existing AGN component \citep[e.g.,][]{2025A&A...702L..17G}, the $\sim$26-fold increase in 1.4~GHz flux density strongly suggests the onset of a new, powerful radio-loud phase. This episode is likely driven by the activation or rapid expansion of a compact jet that now dominates the centimeter emission. 

The overall multi-band behavior characterized by intense radio brightening contrasted with a stable disk--corona dominated X-ray spectrum supports this scenario. Furthermore, the optical--MIR variability appears decoupled from the recent radio outburst, instead reflecting intrinsic accretion fluctuations and subsequent dust reprocessing within the circumnuclear environment.

\subsection{Multi-wavelength Constraints on the Outburst Mechanism} \label{sec:constraints}

The X-ray properties of \src\ provide critical constraints on the nature of the radio outburst. Despite the pronounced radio evolution, the $0.5$--$8$~keV flux remains statistically constant across six epochs spanning 2020--2025. The X-ray spectrum is consistently steep, characterized by a photon index $\Gamma \simeq 3.0$ (Section~\ref{sec:xray}). Such steep continua are a hallmark of radio-quiet NLSy1s and are typically attributed to disk--corona emission or a prominent soft excess \citep[e.g.,][]{Grupe2010}. This contrasts with the harder spectra and rapid variability expected from a jet-dominated high-energy component. The combination of negligible X-ray variability, a steep spectral slope, and intense radio brightening suggests that the new radio-loud phase of a compact jet contributes minimally to the X-ray band, likely due to a high-energy cutoff in the jet spectrum or dilution by the primary accretion flow.

The optical and MIR variability provide a consistent picture of the circumnuclear environment. In the standard AGN paradigm, the dusty torus reprocesses the optical/UV continuum, resulting in MIR light curves that are delayed and smoothed relative to the accretion disk emission \citep{Sheng2017, Lyu2019}. In \src, the long-term fading-then-stabilization pattern observed in the Pan-STARRS $r$- and $i$-bands is mirrored in the \textit{WISE} bands. The MIR emission tracks the optical variability with a time lag of $\tau \lesssim 5$~yr, physically consistent with the light-travel time to the dust sublimation radius for a source of this bolometric luminosity. 

While the precise onset of the radio brightening is unconstrained, the initial VLASS detection in 2017 December is broadly contemporaneous with the MIR light curve entering a stable phase (Figure~\ref{fig:lc}). If the radio emission is physically coupled to the optical--MIR stabilization observed between 2009 and 2014, the radio onset may have occurred as early as $\sim$2010. Synthesizing the multi-wavelength temporal data, we derive an inferred rising timescale of $\tau_{\rm rise} \lesssim 1$~yr, based on the sharp transition of the optical and MIR light curves from a fading to a stabilized state in early 2014.  While the radio onset itself is not precisely sampled, the current radio-bright state suggests the jet may have emerged on a timescale comparable to this circumnuclear transition. This rapid rise is nominally consistent with episodic ejections in black hole X-ray binaries \citep{Fender2004} or blazars \citep{Pietka2015}. However, the protracted stabilization phase ($\tau_{\rm stab} \gtrsim 8$\,yr) significantly exceeds typical transient timescales, potentially indicating the establishment of a sustained, compact jet.

\subsection{Jet Kinematics and Energetics} \label{sec:kinematics}

Modeling the radio SED with a SSA model yields a turnover frequency of $\nu_{\rm p}=0.48\pm0.03$\,GHz and a peak flux density of $S_{\rm p}=38.9\pm4.7$\,mJy. These parameters classify \src\ as a MPS source, characteristic of young jets confined within the host galaxy central environment \citep{Keim2019, Berton2026}. The very flat high-frequency spectral index ($\alpha=0.03\pm0.07$) would imply an unphysically flat optically thin electron index ($p=1-2\alpha\simeq0.94$) if the emission were fully transparent, compared to the canonical p$\sim$2--3 expected from standard diffusive shock acceleration \citep[e.g.,][]{Bell1978}. This favors an interpretation in which the highest observed frequencies remain at least partially self-absorbed or the spectrum is a superposition of multiple compact components with differing turnover frequencies, as expected in inhomogeneous core/jet models that naturally produce flat radio spectra \citep[e.g.,][]{BlandfordKonigl1979}.

Under the assumption of equipartition \citep{Readhead1994}, the SSA results yield an intrinsic angular scale of $\theta \sim 0.62$~mas. At the redshift of \src\ ($z = 0.121$; $1$~mas $\approx 2.19$~pc), this corresponds to a physical radius of $R \sim 0.68$~pc. To estimate the expansion velocity, we define the observed-frame duration of the activity, $\Delta t_{\rm obs} \approx 3.9$\,yr, as the interval between the commencement of the optical/MIR stabilization (mid-2014; see Figure~\ref{fig:lc}) and the initial VLASS detection in late 2017. This yields an average apparent expansion velocity of $\beta_{\rm app} = \frac{R}{c \Delta t_{\rm obs}}(1+z) \approx 0.64$, consistent with typical outflows in radio-loud NLSy1s.

To characterize the compactness and energetic state of the radio emission, we calculate the source-frame brightness temperature using the standard relation:
\begin{equation}
    T_b = 1.22 \times 10^{12} \frac{S_{\rm p}}{\theta^2 \nu_{\rm p}^2} (1+z)  \text{ [K]},
\end{equation}
where $S_{\rm p}$ is the peak flux density in Jy, $\nu_{\rm p}$ is the turnover frequency in GHz, and $\theta$ is the angular size in mas derived from our SSA modeling. Using our best-fit parameters, we find $T_b \approx 6.0 \times 10^{11}$\,K. This significantly exceeds the equipartition limit temperature of $T_{eq} \approx 5 \times 10^{10}$\,K   \citep{Readhead1994}. This discrepancy necessitates a Doppler factor of $\delta = T_b / T_{eq} \approx 12$, implying that the radio emission is strongly boosted.

Furthermore, the expansion velocity $\beta_{\rm app} \approx 0.64$ establishes a minimum bulk Lorentz factor of $\Gamma_{\min} = \sqrt{1 + \beta_{\rm app}^2} \approx 1.19$. However, the high Doppler factor ($\delta \approx 12$) suggests that the actual bulk motion is likely much faster and the jet is viewed at a small angle of $\theta \lesssim 5^\circ$. This geometry is characteristic of the blazar-like orientation commonly identified in $\gamma$-ray-emitting NLSy1s \citep{Abdo2009}.

The derived equipartition magnetic field strength is $B_{\rm eq} \approx 0.25$\,mG, implying an intrinsic jet power of $P_{\rm jet} \approx 1.7 \times 10^{39}$\,erg\,s$^{-1}$. We note that empirical scaling relations for kinetic power based on observed luminosity yield higher estimates ($Q_{\rm jet} \sim 10^{43}$\,erg\,s$^{-1}$). However, this tension is largely reconciled when considering relativistic beaming; de-boosting the observed luminosity ($L_{\rm int} = L_{\rm obs} \delta^{-3}$) brings the kinetic power into closer alignment with our equipartition energy estimates. While these results suggest a significant mechanical output, we emphasize that these energetic derivations remain subject to the inherent uncertainties of radio scaling relations and the foundational assumption of equipartition.

\subsection{Comparison with Jetted NLSy1s and FSRQs} \label{sec:comparison}

In the standard blazar/FSRQ framework, the spectral energy distribution (SED) is governed by the balance between particle acceleration and radiative cooling. More powerful jets typically exhibit lower synchrotron peak frequencies and larger Compton dominance \citep{Fossati1998, Ghisellini1998}, a trend often attributed to enhanced cooling by external photon fields \citep{Sikora1994}. Jetted NLSy1s provide a unique test for this sequence; their high-accretion environments provide substantial broad-line region (BLR) and torus seed-photon fields similar to FSRQs, while their black hole masses and jet powers are generally lower \citep{Komossa2006, Foschini2015}.

The radio-quiet to radio-loud transition in \src\ can be interpreted as the emergence of a new, relativistic jet phase within an NLSy1 accretion environment. The absence of $\gamma$-ray emission in the \textit{Fermi}-LAT band, despite our inferred high Doppler factor ($\delta \approx 12$), provides a key constraint on the nascent jet's SED. While such beaming is characteristic of $\gamma$-ray-emitting NLSy1s \citep{Abdo2009}, the non-detection in \src\ indicates that the inverse Compton peak may be suppressed or located at lower energies. This is consistent with an early-stage jet where the dissipation region remains compact or located deep within the photon field where internal absorption is significant, or where the energy density of external photon fields available for Compton scattering is not yet sufficient to produce GeV emission.

If the jet continues to evolve and the dissipation region moves into a more optimal environment for external Compton scattering, a GeV component may eventually emerge. Conversely, the current lack of a GeV counterpart, despite a high Eddington ratio ($\lambda_{\rm Edd} \approx 0.6$) and a presumably luminous disk/BLR field—points toward a highly compact or nascent jet structure. \src thus provides a rare laboratory to test disk--jet coupling during the earliest stages of jet production. While the transition indicates a significant mechanical output, the precise efficiency of the conversion from accretion power to jet power remains difficult to constrain without direct VLBI monitoring of the jet expansion and component motion. 

\section{Conclusions}\label{sec:conclusions}

We have reported a rare radio-quiet to radio-loud transition in the NLSy1 \src\ and presented a comprehensive multi-wavelength analysis of its emerging jet state. Our main conclusions are as follows:

\begin{itemize}
\item \src\ has entered a persistently radio-bright phase since 2017, driven by a compact jet that now dominates the centimeter-band emission. The observed 1.4\,GHz luminosity ($L_{1.4\,\mathrm{GHz}} \approx 1.33 \times 10^{24}$\,W\,Hz$^{-1}$) represents a 26-fold increase relative to the FIRST-epoch one and significantly exceeds the star-formation threshold.

\item Modeling of the radio SED indicates a turnover frequency $\nu_{\rm p}=0.48\pm0.03$\,GHz, characterizing the source as a megahertz peaked-spectrum (MPS) object. The physical radius inferred from synchrotron self-absorption ($R \sim 0.68$\,pc) and the apparent expansion velocity ($\beta_{\rm app} \approx 0.64$) are consistent with a young, mildly relativistic outflow confined within the circumnuclear environment.

\item The observed brightness temperature $T_b \approx 6.0 \times 10^{11}$\,K necessitates a Doppler factor $\delta \approx 12$, implying a relativistic jet viewed at a restricted angle of $\theta \lesssim 5^\circ$. This orientation aligns \src\ with the blazar-like NLSy1 population, where relativistic beaming significantly enhances the observed flux.


\item Despite the radio outburst, the X-ray spectrum remains steep ($\Gamma \simeq 3.0$) and stable over five years, indicating that the high-energy emission continues to be dominated by the disk--corona rather than the jet.

\item The current radio-loud state (post-2017) coincides with a protracted stabilized phase in the optical and MIR light curves following a decade-long fading trend. While the absence of radio data during the 2000--2017 interval prevents a definitive determination of the radio onset, the multi-wavelength data are consistent with a scenario where the jet emerged during or shortly after the stabilization of the primary accretion flow.
\end{itemize}

High-resolution VLBI imaging and continued multi-wavelength monitoring will be essential to track the evolution of this young jet and further test the disk--jet coupling at high Eddington ratios.

\begin{acknowledgments}
The authors thank the anonymous referee for comments that improved the clarity of this manuscript. This work was supported by the Joint Research Foundation in Astronomy under the cooperative agreement between the National Natural Science Foundation of China (NSFC) and the Chinese Academy of Sciences (CAS) (U1731104), and NSFC grants 12573110, 12433004, 12203014, 12133001, and 11833007. Y.L.Ai. acknowledge the supports from the Shenzhen Science and Technology Program (JCYJ20230807113910021) and the Natural Science Foundation of Top Talent of SZTU(GDRC202208). L.M.D. acknowledges the support from the Key Laboratory for Astronomical Observation and Technology of Guangzhou and the Astronomy Science and Technology Research Laboratory of the Education Department of Guangdong Province.

This work is based on data from the Einstein Probe (EP) mission, a project of the Chinese Academy of Sciences (CAS) in collaboration with ESA, MPE, and CNES. This research has made use of data products from the Wide-field Infrared Survey Explorer (WISE) and the Near-Earth Object Wide-field Infrared Survey Explorer (NEOWISE), which are joint projects of the University of California, Los Angeles, and the Jet Propulsion Laboratory/California Institute of Technology, funded by the National Aeronautics and Space Administration (NASA).

This work utilized the Zwicky Transient Facility (ZTF) and the Asteroid Terrestrial-impact Last Alert System (ATLAS). ZTF is a public-private partnership supported by the NSF and an international collaboration of partners; ATLAS is primarily funded through NASA grants. We also acknowledge observations from LOFAR (the Low Frequency Array) designed and constructed by ASTRON. The Pan-STARRS1 (PS1) data products used in this work were obtained from the Mikulski Archive for Space Telescopes (MAST) at the Space Telescope Science Institute. The PS1 DR2 database accessed is available at \url{https://doi.org/10.17909/s0zgjx37}.
\end{acknowledgments}

\bibliography{j1105}
\bibliographystyle{aasjournalv7}

\end{document}